\documentstyle[aps,prl,twocolumn]{revtex}
\begin{document}
\draft
\preprint{}
\title{ Transport and localization of waves in one-dimensional
 disordered media : Random phase approximation and beyond }
\author{Prabhakar  Pradhan}
\address{Department of Physics,
           Indian Institute of Science,
          Bangalore, 560 012,
          India }
\maketitle
\date{\today}
\begin{abstract}
We report a systematic and detailed numerical study of
statistics of the reflection coefficient $(|R(L)|^2)$
and its associated phase ($\theta$) for a plane wave reflected
from a one-dimensional (1D) disordered medium beyond the random 
phase approximation (RPA) for Gaussian white-noise disorder.
We solve numerically the full Fokker-Planck (FP) equation for
the probability distribution in the ($|R(L)|^2,\theta(L)$)-space
for different lengths of the sample with different "disorder 
strengths". The statistical electronic transport properties
of 1D disordered conductors are calculated using the Landauer
four-probe resistance formula and the FP equation. This
constitutes a complete solution for the reflection statistics 
and many aspects of electron transport in a 1D Gaussian 
white-noise potential. Our calculation shows the contribution
of the phase distribution to the different averages and its
effects on the one-parameter scaling theory of localization. 
\end{abstract}
\vspace{1cm}
\pacs{PACS numbers:\,\, 72.15.-v, 72.15.Rn, 73.23.-b, 05.40.+j}
\section{Introduction}
Wave propagation through 1D random media have been studied for
decades. Typical physical examples of these problems are
electron transport in disordered conductors, light transport in
random dielectric media, etc.. All previous studies agree that
the transmission through a 1D disordered medium decays
exponentially with the length of the sample. For electronic
systems, the conductance decays exponentially with the length of
the sample and the resistance increases exponentially with the
sample length. The main characteristic length scale in the
problem is the localization length. Results of different phases
of research in this field have been reviewed recently in several
review articles~\cite{mott1,mott2,ishii,thouless,abrikosov,erdos,lee}. 

  Disordered 1D systems are quite generic as regards their
transport properties --- disordered 1D conductors are all alike,
but every ordered 1D conductor is ordered in its own way!
Thus, Mott {\sl et. al.} \cite{mott1} first showed that
all the eigenstates of 1D disordered quantum systems are
exponentially localized. There are no truly 1D disordered
metals. For electronic systems quantum interference effects are
important at low temperatures when decoherence due to the
inelastic process may be neglected. At low enough temperatures
with static disorder, the transport properties of the electronic
systems are highly fluctuating (i.e. sample specific) due to the
different conductors with the same impurity concentration 
but with different microscopic arrangements of the impurities,
can differ substantially in their transport properties. Several
studies show that the root mean-squared-fluctuation is more than
the average for length scales larger than the localization
length, while for the length scales within the localization
length (i.e. in the good metallic regime for quasi 1D and
higher dimensional systems) the fluctuations are finite and
universal, the so called universal conductance fluctuation (UCF)
\cite{lee}. These fluctuations make the resistance and the
conductance non-self-averaging quantities. The resistance at low
temperatures when quantum effects are important is non-Ohmic,
i.e., non additive in series. The non-additive nature arises
because of the non-local effects of the quantum wave amplitudes
associated with the electron in the conductor. One has to take
proper care of the phases to add the two quantum resistances
together. 

To calculate any meaningful quantity for a non-self-averaging
quantity, like the resistance and  the conductance, one 
should know the full probability distribution of the same
for every length of the sample. Several authors 
\cite{papa,friesh,kumar,heinrichs,rammal} have derived a
Fokker-Planck equation for the full probability distribution of
the resistance for the Gaussian white noise disorder. Exact
results for the average resistance and the average conductance
can be calculated analytically for the weak-disorder case using
the Fokker-Planck equation. The approximation made in the
weak-disorder case is that the phase of the complex reflection
amplitude relative to the incident wave, or the relative phase
of two non-ohmically additive quantum resistances, are
distributed uniformly, which is the random phase approximation
(RPA). Then the problem can be solved analytically to calculate
the average resistance exactly for the Gaussian white noise
potential. The analytical results show that the resistance and
the conductance have log-normal distributions for large lengths.
The average four-probe resistance increases exponentially with
the length of the sample with a characteristic length scale
called the localization length. The average four-probe
conductance, however, diverges for all length scales. The
divergence of the average conductance for all length scales is
due to resonance (absolute transmission) states which dominate
the probability distribution \cite{lifshitz,azbel}. We will 
discuss elsewhere the issues related to the divergence of the
average conductance. 

It is difficult to get an average quantity by direct analytical
calculation which includes the phase distribution, where the
phase distribution depends on the strength of the disorder.
However, the actual validity of the random phase approximation
has not been studied in detail. There are issues like actual
contribution of the phase distribution to the different
averaging processes that have not been systematically studied so
far. 

We study numerically, the joint probability distribution of the
phase $(\theta)$ and the reflection coefficient ($r$) of the
complex amplitude reflection coefficient ($R(L)=\sqrt{r}
\,exp(i\theta)$) for a one-dimensional disordered conductor with 
the Gaussian white noise disorder, for different lengths of the
sample and different strengths of the disorder. This is mainly
the solution of the Fokker-Planck equation, where the
probability density is varying in the $(r,\,\theta)$-space and 
evolving with the sample length.  Using the "invariant imbedding"
technique, a non-linear Langevin equation can be derived for
the $R(L)$. Then, using the stochastic Liouville equation for
the probability evolution and the Novikov theorem to integrate
out the stochastic aspect due to the random potential, one gets
the Fokker-Planck (FP) equation in the $(r,\theta)$-space for
varying sample length $L$ \cite{kumar,rammal}. Integrating
$\theta$ and $r$ parts separately from the joint probability
distribution $P(r,\theta)$, we have calculated the marginal
probability distributions for $r$ and $\theta$, respectively,
for a given length and disorder strength of the sample. The
Landauer formula \cite{landauer,anderson} relates the resistance
and the conductance directly to the reflection coefficient (r).
Once the reflection probability is known, the average electronic
properties of the system such as resistance, conductance $etc.$
can be calculated easily using the Landauer formula. 
We have outlined the range of validity of the random phase
approximation, for the parameters on which the phase
distribution depends, namely the localization length and the
wave vector of the incoming wave. Our calculation shows that
$\xi\geq\lambda=2\pi/k$ is the limit where the random phase
approximation is valid, where $\xi$ is the localization length
and $\lambda$ is the wave length of the incoming wave (and for
the electronic case we will consider $\lambda$  to be Fermi wave
length)

Knowing the probability distribution of the reflection
coefficient and using the Landauer four-probe
resistance/conductance formula, we have calculated the
probability distribution for the resistance/conductance 
and find it to be log-normal, for all strengths of the white
noise Gaussian disorder. We calculate the averages and the
fluctuations of the reflection coefficient, the conductance and
the resistance and of logarithms of the conductance and the
resistance. Lastly, we study the effects of the phase
distribution on the one-parameter scaling theory of
localization. We have shown that the one-parameter scaling
theory holds quite well for large sample lengths, but the phase
has a definite but small effect on it. To the best of our
knowledge this is the first work where the joint probability
distribution for the reflection coefficient and its phase for a
1D Gaussian white-noise disordered conductor is calculated for
all length scales and for different disorder strengths. The way
we have calculated the different quantities here involves
essentially no approximations. We will discuss works of others
while discussing our results. We report here mainly the results
of the electronic transport in disorder conductors, but some
parts of the formalism apply equally well to the case of light
transport in random dielectric media. 
\section{Method of calculation} 
Both the Schr\"{o}dinger and the Maxwell equations can be
transformed to the Helmholtz equation. Therefore, these two
equations can be studied in the same frame work for the
reflection and the transmission amplitude, since they transform
to the same Helmholtz equation and have similar initial
conditions. 

Consider the Schr\"{o}dinger Equation,
\begin{equation}
- \frac{\partial^2 \psi}{\partial x^2} + V(x)  \psi
\, = k^2 \psi,
\label{sceq}
\end{equation}
where we have set $\hbar^2/2m =1 $, \\
and the Maxwell equation,
\begin{equation}
\frac{\partial^2 E}{\partial x^2} + \frac{\omega ^2}{c^2}
[\epsilon_0 + \epsilon (x)] E
\, = 0 \,.
\label{mxeq}
\end{equation}
Transforming the Schr\"{o}dinger and the Maxwell equation
to the standard Helmholtz equation form, we obtain :
\begin{equation}
\frac{\partial^2 u}{\partial x^2} +  k^2 \left[ 1+\eta(x) \right] u
\, = 0,
\label{hmeq}
\end{equation}
      where  $\eta(x)$=$-V(x)/k^2$ for Schr\"{o}dinger  wave, and
$\eta(x)$= $+{\epsilon(x)\over  
\epsilon_0 } $ and $k^2={\omega^2\over c^2}\epsilon_0$ for the Maxwell
wave, where $\epsilon_0 $ is the constant dielectric background
and $\epsilon(x)$ is the spatially fluctuating part of the dielectric
medium. 
\subsection{The Langevin Equation for the complex 
amplitude reflection coefficient R(L)}  
Consider a plane wave of wave vector $k$ incident from the
right side of the disordered sample of length $L$ having the
reflection amplitude $R(L)$. The  non-linear Langevin equation
for the reflection amplitude for the plane wave scattering problem
can be derived by the invariant imbedding technique \cite{rammal}. 
The Langevin equation for the complex amplitude reflection
coefficient is : 
 \begin{equation}
  \frac{dR(L)}{dL} \, = \, 2ik R(L)\, + \,i \frac{k}{2}
  \eta(L)(1+R(L))^2,
\label{leq}
\end{equation}
with the initial condition R(L)=0 for L=0.

The main idea is to use the Langevin equation to get a FP
equation for the reflection probability density. For the electronic
system consider the reflection coefficient (or the transmission
coefficient) of a free electron wave scattered by the random
potential in order to calculate the resistance/ conductance of
the system.  The average properties of the
resistance/conductance can be calculated using the Landauer
four-probe formula and the FP equation for the reflection 
coefficient. 
\subsection{The Fokker-Planck equation}
To get the Fokker-Planck equation from the non-linear Langevin
equation \cite{rammal}, one has to get the probability density
equation first and then the stochastic aspect due to the random
potential has to be integrated out. The Langevin equation
(Eq.\ref{leq}) can be solved analytically for the Gaussian white
noise potential to get the FP equation. The non-linear Langevin
equation for $R(L)$ in (Eq.\ref{leq}) is basically two coupled
differential equations for the magnitude and the associated
phase parts.\\ 
Taking
\begin{equation}
 R(L) = \sqrt{r(L)} exp(i \theta(L)),
\label{rap}
\end{equation}
and substituting Eq.\ref{rap} into Eq.\ref{leq}, and  equating 
the real and the imaginary parts on both sides of the Eq.\ref{leq},
one gets, 
\begin{equation}
\frac{d r}{d L}
\, =  k \eta (L) r^{\frac{1}{2}}(1-r)\sin\theta,
\label{tceq1}
\end{equation}
\begin{equation}
\frac{d \theta}{d L} =
2 k + {k\over 2} \eta(L)\left[ 2 + \cos\theta ( r^{1/2} + r^{-1/2} )\right].
\label{tceq2}
\end{equation}
Now, according to the van-Kampen lemma \cite{van}, these two 
stochastic coupled differential equations will produce a 
flow of the density $Q(r,\theta)$ in the $(r,\theta)$-space
according to the stochastic Liouville equation with 
increasing length of the sample, i.e., $Q(r,\theta)$ is the
solution of the stochastic Liouville equation:
\begin{equation}
\frac{\partial Q(r,\theta,L)}{\partial L} =
- \frac{ \partial }{\partial r}
 ( Q \frac {d r}{d L}) 
-  \frac{\partial}{\partial \theta}  
 ( Q  \frac{d \theta}{d L} ) ,
\label{lveq1}
\end{equation}
where $dr/dL$ and $d \theta/dL$ are given by
Eqs.\ref{tceq1} and \ref{tceq2}. Now, substituting the values of $
dr/dL$ and $d\theta/dL$ in the Eq.\ref{lveq1} one gets,
\begin{eqnarray}
\frac{\partial Q}{\partial L} &=&
  -k \sin\theta \frac{d}{dr} \left[r^{1/2} (1-r) \eta(L) Q\right]
- k \frac{\partial}{\partial \theta} \left[ \eta(L) Q \right] \nonumber\\ 
&& - 2 k \frac{\partial Q}{\partial \theta}
   - {k\over 2} (r^{1/2} + r^{-1/2}) \frac{\partial}{\partial \theta} 
     \left[ \cos\theta \eta(L) Q  \right].
\label{lveq2}
\end{eqnarray}
To get the Fokker-Planck equation, Eq.\ref{lveq2} has to 
be averaged over the stochastic aspect, i.e., over all realizations 
of the random potential.  For the case of a Gaussian white noise
potential, Eq.\ref{lveq2} can be averaged out over the
stochastic potential analytically using Novikov's \cite{novikov}
theorem. For the Gaussian white noise potential: 
\begin{equation}
< \eta (L) > = 0\,\,\,and\,\,\, < \eta(L) \eta(L') > = q \delta (L-L') .
\label{geq}
\end{equation}
Eq.(\ref{lveq2}) has terms like $\eta Q $ which are to be
averaged out. 
For the Gaussian white noise
disorder, the Novikov theorem says :
\begin{equation}
<\eta(L) Q[\eta] >_\eta = {q\over 2} <\frac{\delta Q[\eta] }{ \delta
\eta(L)} >.
\label{nkt}
\end{equation}
After averaging out the disorder aspect in Eq.\ref{lveq2} and writing
$<Q(r,\theta)>_{\eta} \equiv P(r,\theta)$, one gets the Fokker-Planck
equation: 
\begin{eqnarray}
\frac{ \partial P(r,\theta) }{\partial l} 
&=& \left[ \sin \theta \frac{\partial}{\partial r} r^{1/2} (1-r)
    + \frac{\partial }{\partial \theta} \right. \nonumber \\
&+& \left. {1\over 2} (r^{1/2} + r^{-1/2})
\frac{\partial}{\partial \theta} \cos\theta \right]^2
P(r,\theta) \nonumber\\
&-& 2k\xi{\partial P(r,\theta) \over \partial \theta}.
\label{fpeq}
\end{eqnarray}
Where $l\equiv L/\xi$ and $\xi\equiv ({1\over 2}qk^2)^{-1}$ is the
localization length. 

The Fokker-Planck equation, Eq.\ref{fpeq} has all the information
of the probability distribution of the reflection coefficient
$(r)$ and the associated phase ($\theta$) for different length
scales of the sample and with varying disorder strengths. We
will explore several aspects of the Eq\ref{fpeq} in our study,
which we are going to discuss in detail in later Sections. 
\section{Parameters of the problem}
The Fokker-Planck equation (Eq.\ref{fpeq}) has three
parameters to describe the problem fully:
(1) the length of the sample $ L$, (2) the localization
length  $\xi$, and (3) the incident wave vector $k$.
In the re-arranged form of the Eq.\ref{fpeq}, as it is written, 
it has effectively two parameters: $l= L/\xi$ and $C= 2k\xi$. 
Here $l$ is a number which gives the length  of the sample
in units of the localization length and $C$ is a number which
fixes inverse of the disorder strength in terms of the wave
vector of the incident wave and the localization length.
Larger value of $C$ implies that the localization length is large,
or the incoming electron energy is higher, or both, that is, the
weak-disordered regime. Conversely, when $C$ is small it means
$\xi$ is small, or the incoming wave energy is small or both, 
that is, the strong disorder regime.
\section{Analytical solutions for the resistance and the conductance
in the random phase approximation (RPA)}
\subsection{Resistance}
In the random phase approximation (RPA), which is valid for weak
disorder and large incident electron energies, one can write
$P(r, \theta ) = (1/2\pi) P(r)$ i.e. $P(r,\theta)$ factorizes,
and $\theta$ is uniformly distributed over $2\pi$.
Considering $\partial P/\partial \theta =0$, the Fokker-Planck
equation Eq.\ref{fpeq} in $r$ then becomes:
\begin{equation}
\frac{\partial P(r)}{\partial l} = \frac{\partial }{\partial r}
\left[ r\frac{\partial }{\partial r}(1-r)^2 P(r) \right]. 
\label{fpeqr-rpa}
\end{equation}   
( Here we have used the same symbol $P(r)$ for the marginal
probability density of $r$ as for the joint density
$P(r,\theta )$ .)

Now, from the Landauer four-probe resistance formula, the
dimensionless resistance as a function of the reflection coefficient
is: 
\begin{equation}
 \rho(l)= \frac{r(l)}{1-r(l)}.
\label{land}
\end{equation}
Making the transformation from $P(r)$ to $P(\rho)$ by using
$P(\rho)=P(r)(dr/d\rho) $, one obtains the Fokker-Planck equation
for the resistance: 
\begin{equation}
\frac{\partial P(\rho)}{\partial l} = \frac{\partial }{\partial
\rho}\rho(1+\rho) \frac{\partial }{\partial \rho} P(\rho) 
\label{fpeq-rho},
\end{equation}   
with the initial condition, $P(\rho)=\delta(\rho)$ for $l=0$.

Eq.\ref{fpeq-rho} has been derived earlier by several authors
\cite{melnikov,kumar,rammal}. This equation has also been
derived from the maximum entropy principle (MEP) by Mello and
Kumar \cite{mello}. 

The average of $\rho^n$ can be obtained without solving 
directly Eq.\ref{fpeq-rho} as following.\\
Let us define,
\begin{equation}
\rho_n = \int_0^\infty P(\rho) \rho^n d\rho
\label{rec-eq}
\end{equation}
Multiplying Eq.\ref{fpeq-rho} by $\rho^n$ and integrating both
sides of the equation for $\rho$ from $0$ to $\infty$, one gets
a moment recursion equation for the average moments of the
resistance, 
\begin{equation}
{d\rho_n \over dl} = n(n+1) \rho_n + n^2 \rho_{n-1}.
\label{recor}
\end{equation}
Since the probability is always normalizable, the value of
$\rho_0$ will be:
\begin{equation}
\rho_0 = \int_0^\infty P(\rho) \rho^0 =\int_0^\infty P(\rho) =1
\label{pnorm}
\end{equation}
Once we know the initial value $\rho_0$, then Eq.\ref{recor} can
be solved analytically for higher moments of the average
resistance,
\begin{eqnarray}
 \rho_1(l)&=& {1\over 2} (e^{2l} - 1), \\
 \rho_2(l)&=& {1\over 12}(2e^{6l}-6e^{2l} + 4), \\
 \rho_n(l)& \simeq & {e^{n(n+1)l}\over n(n+1)!}, \,\,\,n\gg 1\,\,
\end{eqnarray}
 and specifically,
\begin{equation}
\mbox{ rms \,\,fluctuation \,\,of}\,\,\, \rho \sim  e^{3l}.
\label{fluc}
\end{equation}
The above expressions imply that the average resistance increases
exponentially with the length of the sample and the rms fluctuation
is greater than the average, indicating that the resistance is not a 
self-averaging quantity. It is clear now why one has to consider
the full probability distribution to describe the disordered
quantum resistors. Eq.\ref{fpeq-rho} has also an analytical
expression for the full distribution of $\rho$ for the large $l$
limit \cite{kumar}, which is  log-normal.
\subsection{Conductance} 
If one solves the Fokker-Planck equation for  $<g^n>
(=<1/\rho^n >)$, it has been shown by several workers and
explicitly by Melnikov \cite{melnikov} that all the moments
$\nu$ of $<g^\nu>$ diverge for $\nu > 1/2 $, for all length
scales of the sample. The cause is the existence of the resonant
states, i.e., states $r\sim 0$. Though, the probability is less
for the resonant states, but it is finite. The existence of the
resonance states have been discussed in detail in Ref.
\cite{lifshitz,azbel}. 
\section{Aim of our work}
The main aim of our work is to evolve the full Fokker-Planck
equation, (Eq.\ref{fpeq}), numerically. Eq.\ref{fpeq} has full
informations of the the statistics of the reflection coefficient
and its phase for the case of Gaussian white noise disordered
potential. We will study several aspects of the solution, and
try to see the actual contribution of the phase, which goes
beyond the random phase approximation. Important point to note
here is that the numerical calculation part involves only the
solution of the Fokker-Planck equation which has been derived 
analytically without approximation. 
\section{Numerical details}
We took $r$ and $\theta$ as Cartesian variables in
a two-dimensional $50\times 50$ grids. An explicit
finite-difference scheme \cite{nrp,ame} was used to solve the FP
equation. The von-Neumann stability criterion was checked and
the Courant condition for the used discrete iterative length was
strictly maintained. A few results were also checked by using
the rather time consuming implicit finite-difference scheme.
\subsection{Allowed error bars}
Error bars of the order of $10^{-4}$ for $r$, $3\times 10^{-3}$
for $\theta$, and $10^{-12}$ for length $l$, were allowed for 
the whole range of numerical calculations.
\subsection{Initial probability distribution $P(r,\theta)$ at $L=0$} 
The FP equation (\ref{fpeq}) poses an initial value problem. The
initial probability distribution $P(r,\theta)$ at $l=0$ has to be
specified, which will then evolve with the increase of the length of
the sample. The Fokker-Planck equation (\ref{fpeq}) is however
singular at r=0. This causes a technical problem for solving the
equation numerically. To circumvent this problem, we have
therefore taken an initial (fixed) scatterer with $r=.01$ by
putting a half-delta function potential peaked at $l=0$ which
could be physically understood as due to an initial impurity
sitting at $l=0$, or the contact resistance of the leads. By
"fixed" we mean that it is fixed over all the realizations of
the sample randomness. Phase distribution for such a weak
delta-function potential will peak around $+\pi /2$ or $ -\pi /2
$, depending on the sign of the delta-function potential. 
Once $r_0=.01$ is fixed, then the phase distribution has equal
probability peak at $+\pi/2$ and $-\pi/2$. A fixed weak delta
scatterer at the position $l=0$ will not change the gross
statistics, except at very smaller length scales. We have kept
this initial distribution same throughout the numerical
calculations. We could not consider any smaller value of the
initial-fixed-reflection coefficient ($r_0$), or a lower cut-off
to the $r=0$ singularity, for reason of convergence criterion of 
the numerical algorithm. An estimate can be done for the initial
cut-off length $l_0$ of the sample (in terms of the localization
length $\xi$) for this small $r=r_0$. Taking analytical results
for the weak disordered case, one gets: 
\begin{equation}
   \rho_0 = {r_0\over (1-r_0)} =  {1\over 2}[exp(2l_0)-1]
\end{equation}
then
\begin{equation}
  l_0= L_0/\xi= 2ln[{1+r_0\over 1-r_0}] 
\simeq {1\over 2}ln(1+2r_0)\simeq r_0=.01 
\label{int-length}
\end{equation}
This implies that the initial length is $1\%$ of the
localization length, throughout the numerical calculation.

For numerical calculation, the delta-function has to be taken as
the limit of a continuous function. In $(r,\theta)$-space for a
physically reasonable initial probability distribution, we have
taken this as 
\begin{equation}
  P(r,\theta )_ {l=0}  = \delta (r -.01) \left[\delta (\theta - \pi /2)
   + \delta (\theta + \pi /2) \right],
\label{in-dis}
\end{equation}
where the delta functions are sharp Gaussians.

Fig.\ref{pint} shows this initial probability distribution 
$P(r,\theta )$ nominally at $l=0$, and the marginal distribution $P(r)$
of the reflection coefficient and the marginal distribution
$P(\theta)$ of the phase $\theta$. It should be emphasized again
here that the initial probability distribution of the phase
${\theta}$ can be taken at any small enough length. However, the
statistical properties of the system do not depend on the
initial distribution except for very small lengths. 
\subsection{Boundary conditions for $r$ and $\theta$ for any
length $L$}
The unit step length of the discrete evolution is taken to 
be $\triangle l = 10^{-6}$. For every discrete evolution, we
took $P(r, \theta)=0$ for $r >1$ and $r<0$ along $r$ axis; and 
the boundary condition was taken as periodic along $\theta$
axis, such that $P(r,2\pi +\triangle \theta) =P(r,\triangle
\theta)$ for every discrete evolution. 
\section{Results and discussions}
\subsection{Evolution of {$P(\protect{r},\theta )$} with the length
\protect{$L$} for different  disorder strengths }
We will consider the evolution of the full probability
distribution $P(r,\theta )$ for different lengths $l$ for the
three main regimes of disorder strength ---(1) weak, (2) medium,
and (3) strong. 
\subsubsection{ $P(r,\theta)$ for weak disorder : $2k\xi=100$ 
$(2k\xi >>1$)}
Figs.\ref{pevw}(a),(b),(c),(d) show typical distributions of the
probability in $(r,\theta)$-space for different lengths $l=1,\,
2,\, 5,\,\mbox{and} \,10$. Evolution shows clearly that the
phase part is uniformly distributed. Evolution of the
probability is mainly along the amplitude ($r$) axis.
Probability evolves with length and peaks near $r=1$ for large 
lengths. These evolution pictures imply that the random
(uniform) phase distribution holds for the weak disorder, or for
the large values of the localization length. Later we show the
nearly exact range for the disorder parameter ($2k\xi$) where the
random phase approximation is valid.
\subsubsection{ $P(r,\theta )$ for medium disorder:
 $2k\xi=1 $ ($2k\xi\sim 1 $)}
Similarly, Figs.\ref{pevm} (a),(b),(c),(d) are as Fig.\ref{pevw}
but for the disorder parameter $2k\xi=1$, i.e., medium disorder
regime. 
These evolutions clearly show that the probability
distribution $P(r,\theta)$ is quite complicated and
has an asymmetric distribution in phase. The phase distribution 
peaks on one side $(\theta > \pi)$ for large lengths. 
\subsubsection{ $P(r,\theta)$ for strong disorder : $2k\xi=.001$
$(2k\xi<<1)$} 
Similarly, Figs.\ref{pevs}(a),(b),(c),(d) are as Fig.\ref{pevw},
but for the disorder parameter $2k\xi=.001$, i.e., the strong
disorder regime. 
Probability distribution is perfectly symmetric centered at the
phase $\theta = \pi$. In the strong disorder parameter regime the
sample tries to behave as a perfect reflector and totally
reflect back the incoming wave with opposite phase. Evolution
pictures show that $P(r,\theta )$ peaks around $r=1$ and is
symmetrically around $\theta=\pi$ for larger lengths. It will be
shown later that the distribution does not change substantially
with further increase of the disorder strength parameter, i.e.
the distribution is insensitive to the disorder parameter
$2k\xi$ in that regime. 
\subsection{ Marginal distribution $P(r)$ of the reflection coefficient
($r$) and Marginal distribution $P(\theta)$ of the phase ($\theta$) for
strong, medium and weak disorders} 
Figs.\ref{marrt}(a),(b),(c) show $(i)$ the marginal probability
distribution $P(r)$ when the phase($\theta$) part of the
$P(r,\theta)$ is integrated out and $(ii)$ corresponding
marginal distributions $P(\theta)$ of the phase $\theta$ when
$r$ part of the full distribution $P(r,\theta)$ is integrated
out for the three cases of disorder regimes considered
above are \,\,\,\,\,the  (a) weak , (b) medium and (c) strong. 

 Fig.\ref{marrt} clearly indicate that the phase $(\theta)$
distributions are very different for the three different regimes
of disorder parameters considered here. However, the reflection
coefficient $(r)$ distribution does not show much variation with
the strength of the disorder parameter. In fact, they are quite
similar. We will emphasize this point again when we later
discuss the one-parameter scaling theory. The behavior of the
probability distribution indicates that the distribution of the
phase has little effect on the distribution of the reflection
coefficient $P(r)$, and hence on the average transport
properties. 
\subsection{Distribution of the phase ($\theta$) with the
strength of disorder parameter $2k\xi$}
We now study the phase distribution $P(\theta)$ with the
strength of the disorder parameter ($2k\xi$). We consider here
mainly two cases of fixed lengths: (1) $l=1$, and (2) the
asymptotic limit when $l\rightarrow \infty$. 
\subsubsection{ The phase $(\theta)$ distribution $P(\theta)$ 
 for different disorder parameters $(2k\xi)$  for fixed
length $l=1$.}        
Fig.\ref{ptl1} shows the plot of $P(\theta)$ for the fixed
length $l=1$ , with different values of the disorder parameters
$(2k\xi)$. Figures show that in the strong as well as
in the weak disorder limit $P(\theta)$ is insensitive to the
strength of the disorder parameter. Phase distribution is
uniform for the weak disorder case. It has a perfectly symmetric
peak centered at $\theta=\pi$ for the strong disorder case.
In the case of intermediate disorder, the distributions are
asymmetric with respect to the $\theta=\pi$ point. For the
medium disorder strength parameter, $P(\theta)$ interpolates
continuously between the strong and the weak disorder limiting
phase distributions. It is clear from the $P(\theta)$
distributions, that the random phase approximation is valid for 
the condition $ \xi > \lambda=2\pi/k $, where $ \lambda $ is the
wave length of the incoming wave and $\xi$ is the localization
length. (This has been checked systematically in our numerical
calculation.) 
\subsubsection{The phase $(\theta)$ distribution $P(\theta)$ in
the asymptotic limit of large lengths} 
Solution of the FP equation (Eq.\ref{fpeq}) gives the full
probability distribution $P(r,\theta)$ in the $(r,\theta
)$-space. For larger lengths, one can make the approximation:
$r\approx 1$ and $P(r)\approx\delta (r-1)$. In this asymptotic
limit, marginal distribution for phase $P(\theta)$ becomes:
\begin{equation}
P(\theta) = \int_0^1 P(r, \theta ) \delta (r-1) dr \equiv P(1,\theta).
\label{asymp-r}
\end{equation}
Now from Eq.\ref{fpeq} and Eq.\ref{asymp-r} we get:
\begin{eqnarray}
\frac{\partial P(\theta)}{\partial l}
 &=& \frac{\partial }{\partial \theta}\left[(1+\cos\theta)
\frac{\partial }{\partial \theta}(1+\cos\theta) \right]
P(\theta) \nonumber \\
&-& 2k\xi \frac{\partial P(\theta )}{\partial \theta}.
\label{asymp-phase}
\end{eqnarray}
Fig.\ref{ptl2} shows the plot of $P(\theta) $ in the asymptotic 
limit for different disorder strength parameters ($2k\xi$),
as the parameter varies from the weak to the strong disorder
limits. Pictures show that in the interval from $0$ to $2\pi$ a
symmetric double peaked distribution around $\pi$ in the strong
disorder regime; a uniform phase distribution in the weak
disorder regime; and an asymmetric phase distribution in the
intermediate disorder regime. These distributions indicate that
the essential features of $P(\theta)$ have not changed with the
$r\approx 1$ approximation, relative to the $l=1$ case which is
shown in Fig.\ref{ptl1}. 
\subsection{ Discussions on the phase distribution }
We saw that the random phase approximation (RPA) is valid for
$\xi > \lambda$. Physical meaning of the RPA is that the
incoming wave has to undergo multiple reflections before
escaping a localization length and in the process the wave
randomizes its phase. For the weak disorder case the
localization length is large. Hence the phase of the wave gets
randomized. In the other extreme case of the strong disorder,
system tries to behave as a perfect reflector. Hence, phase of
the reflected wave tries to peak at $\pi$ (opposite phase with
respect to the incoming wave). In the regime of intermediate
disorder, $P(\theta)$ distribution is disorder specific and has
some bias points to peak. 

At this point let us discuss the results of other workers. Phase
distribution for 1D systems have been studied earlier by Sulem
\cite{sulem}, Stone {\sl et. al.} \cite{stone}, Jayannavar
\cite{jayan}, Heinrichs \cite{hein} and Manna {\sl et. al.}
\cite{manna}. Sulem missed the random phase distribution in the
limit of weak disorder, and his calculated phase distributions
show peak near $\pm \pi$, and it does not look like a symmetric
distribution. The study of Stone {\sl et. al.} showed for the 1D
Anderson model that (i) for the case of weak disorder the phase
distribution is uniform, (ii) non-uniform for the strong
disorder, (iii) and pinning of phase distribution near $2\pi$
for very strong disorder and large lengths. 
Their works also confirm that the distribution of the phase is
insensitive to the disorder strength in the limit of weak as
well as strong disorder limits. Manna {\sl et. al.} show uniform
phase distribution from 0 to $2\pi$ for the weak disorder, 
peaking of the phase distribution near $\pi$ for the strong
disorder. Jayannavar's calculation for the asymptotic
$(l\rightarrow \infty)$ phase distribution shows a uniform
distribution of phase for weak disorder. However, the phase
distribution for strong disorder shows several peaks that do not
agree with our results, though we have numerically solved the
same FP equation. 
\subsection{ Marginal probability distribution $P(r)$ of the
reflection coefficient $r$ with respect to the disorder strength
parameter $2k\xi$} 
Fig.\ref{prl1}  shows the probability distribution of the
reflection coefficient with the disorder parameter strength
$(2k\xi)$ for the sample length $l=1$. It shows very little
dependence on the strength of disorder parameter. But the
distribution certainly has a small spread. 

We can see that though the phase $(\theta)$ distributions
$P(\theta)$ is quite different for different strengths of the
disorder parameter $(2k\xi)$, the reflection coefficient $(r)$
distribution $P(r)$ does not change appreciably with the
disorder strength for a fixed length of the sample.
\subsection{Probability distributions for: $g$, $\rho$, $ln(g)$
and $ln(\rho )$ } 
Once the marginal distribution $P(r)$ is known, the probability
distribution of any quantity which is a function of $r$, can be
easily calculated through the Jacobian of the probability
transformation.
Thus, from the Landauer four-probe resistance formula
Eq.(\ref{land}) one can easily calculate the probability
distributions for the resistance($\rho$), conductance($g$),
$ln\rho$ and $ln g$ , by multiplying with the proper 
Jacobian in to $P(r)$. We can write:
\begin{eqnarray}
         P(\rho )&=& P(r) (1-r)^2,  \\
         P(g)   &=& P(r) r^2 ,      \\
         P(ln(\rho ))&=& P(r) r (1-r), \\
         P(-ln (g))  &=& P(r) r (1-r). 
\label{}
\end{eqnarray}
(Here we have used the same symbol $P$ for before and after the
transformation.)\\
Fig.\ref{prho} shows the plot of  $P(\rho$ ), and  \\
Fig.\ref{pcon} shows the plot of $P(g)$, for the a fixed
length $l=10$, and for the three different regimes of disorder.
These distributions show a log-normal form for all disorder
strengths. 

Fig.\ref{pln} shows plots of $P(ln(\rho)) (\equiv P(-ln(g)))$
vs $ln(\rho)$ ($\equiv -ln(g))$, for the fixed length $l=10$. 
These distributions are Gaussian. (Probability distribution for
the $ln(\rho)$ or $-ln(g)$ being a Gaussian implies that the
$\rho$ and $g$ obey a log-normal distribution.)

Fig.\ref{plnmd} shows the $ln(\rho )(\equiv -ln(g))$
distribution for the medium disorder with disorder parameter
$2k\xi=1$, for different lengths $l$. The distribution matches
with the Gaussian form quite well even for the case of $l=1$. 

We can conclude that the probability distributions for $\rho$
and $g$ are log-normal even for a short sample length ($\sim$ 
localization length). They show little dependence on the
strength of the disorder. The nature of the distribution is not
affected by the phase fluctuations. 
\subsection{Mean and rms fluctuation of $r$, $g$ and $\rho$ }
Figs.\ref{average}(a),(b),(c), show the plots of the averages and the
root-mean-squared fluctuations of $r$, $\rho$ and $g$ against
the sample length $l$ for different disorder strength parameters.
From the figures one can see $<r>$ first increases rapidly and
then saturates slowly to $r=1$ for the large lengths. The
average resistance ($<\rho>$) shows exponential increase and the
average conductance ($<g>$) shows an exponential decrease with
the sample length 
$l$. The fluctuations in the case of $\rho$
and $g$ are always more than the average as shown in the figure.
However, in case of $r$ the fluctuations are always less than
the average and are finite. 
At this point we would like to mention that the average
quantities like average resistance do not admit proper treatment
by this type of numerical calculation for large length. This is
because the Landauer four-probe resistance formula is
$\rho={r\over 1-r}$, that is, $\rho$ is singular at r=1. The
$P(r)$ never saturates (i.e., no steady state solution), but is
localized near $r=1$ for large lengths. For larger lengths,
contribution to $<\rho>$ is dominated by the $r$-values very near
to $r=1$ ($\equiv \rho\rightarrow\infty$) singularity, and 
numerically it is very difficult to include this long-tail
$\rho-distribution$ for the resistance.
However, numerical calculation of the probability distribution
of a quantity which is a non-singular function of $r$,
does not pose much problem.

\subsection{ One-parameter scaling theory and the phase }
Here we are re-examining the effect of disorder strength
and phase distribution on the one-parameter scaling theory of
localization. The scaling theory has been discussed in
Ref.\cite{gfour,agfour}. According to the ansatz of the scaling
theory, the dimension-less conductance $g$ (in units of $e^2/\pi
\hbar$) obeys a universal scaling relation, 
\begin{equation}
{d ln(g) \over d ln(l)} =\beta(g).
\label{}
\end{equation}
The above scaling function $\beta(g)$  depends only on
the dimensionality. The scaling relation suggest that the
fractional change of the conductance per unit fractional change
of the length depends only on $g$ at that length; that is, once
$g$ is known at any length scale, one can derive the conductance
for all lengths. The central idea of the scaling theory lies in
the definition of the $\beta$-function which is assumed to have
a smooth one-parameter scaling behavior with the system 
size. Now, it is well established that these assumptions are
in general not correct because the scaling theory neglects 
fluctuations. For the case of pure elastic scattering, the
resistance/conductance is not a self-averaging quantity.
Fluctuations increase with system size for insulating regime,
and in the metallic regime there is UCF, which has a finite
non-zero value. These issues have been debated by several 
authors \cite{efetov,ioffe,kumarcon,alt}. \\
The points are the following:\\
(1) The precise meaning of the scaling function $\beta(g)$ is
not very clear,\\ 
(2) $g$ is a random quantity which depends on the microscopic
details of the impurities inside the localization length,
and only for length scales larger than the localization length
$g$ may have some universal properties. Only defining $g$,
independent of the microscopic details, will not specify the
average behavior of the sample.\\ 
The question then is how to define a scaling function which can
accommodate fluctuation effects too. It is meaningful to define
statistical properties like the probability distribution of the
resistance/conductance in terms of which one can calculate
the average properties of the system. One can then look for the
number (one,or more) of parameters needed to specify the
probability distribution  \cite{anderson,shapiro1}. 

There have been several studies
\cite{kumar,anderson,shapiro1,shapiro2} that show that for 
1D in the weak disorder limit (i.e. within RPA) , $ln(\rho)$ has
a normal distribution (we also have discussed the same in the
earlier Section), in the limit of large lengths. 

Now, the expression for the Gaussian distribution for logarithm
of the resistance will be: 
\begin{equation}
P(\ln\rho) = {1\over\sqrt{2\pi\sigma^2}}
 exp\left[{-(ln\rho-<ln\rho>)^2\over 2\sigma^2} \right]
\label{logdis}
\end{equation}
which requires two parameters: the mean $< ln\rho>$ and the
standard deviation $\sigma$. 

It was found \cite{slevin} that in the random phase approximation (RPA)
the $n$th cumulant $C_n$ for the large lengths $l$,
\begin{equation}
C_n(\ln\rho ) \rightarrow l a_n,
\label{}
\end{equation}
where $a_n$ is some constant.
It has also been seen by an exact solution for the weak disorder
case \cite{kumar} that,
\begin{equation}
  C_1=a_1 l \,\,\,\,\,\,\mbox{and} \,\,\,\,\,\, C_2=a_2 l
\label{}
\end{equation}
Eq. (\ref{logdis}) can now be rewritten as:
\begin{equation}
P(\ln\rho) = {1\over\sqrt{4\pi a_2 l}} 
 exp\left[{-(ln\rho-  a_1 l)^2 \over 2 a_2 l} \right]
\label{}
\end{equation}
The above Gaussian distribution has clearly two parameters ($a1$
and $a2$) but can be reduced to a single parameter if the
average and the rms fluctuations (variance) are universally
related. 
Knowing the probability distributions $P(r)$ for different lengths
and disorder strengths, we can check the validity of the
one-parameter scaling theory for the probability distributions 
$P(ln\rho)$.

In Fig.\ref{pone} we have plotted  $var \,ln\rho$  vs  $<ln\rho>$ 
for the cases of(1) weak, (2) medium, and (3) strong disorder.
To obtain these plots we have calculated $<ln\rho>$ and
$var$  $(ln\rho)$, for a chosen value of the disorder parameter 
$2k\xi$, as function of the sample length. Different plots 
thus correspond to the different value of the disorder parameter
strengths. A One-parameter scaling would have these plots coincide.
The figure shows that the $var$  $(ln\rho)$ vs $<ln\rho>$ plots
follow approximately the same graph for the different disorder
strengths and nearly coincide for large lengths. To this extent
we have a deviation from the one-parameter scaling theory in 1D.

\section{ Discussion and Conclusions }
We have solved here the 1D transport problem for the case of Gaussian
white-noise disorder numerically.
This is a nearly complete solution of transport properties of
the 1D Gaussian white-noise random potential that goes beyond
the conventional random phase approximation (RPA) valid only for 
weak disorder. We have evolved the full probability distribution
in the reflection coefficient($r$) and the associated phase
$(\theta)$ space (i.e. ($r,\theta$)-space) of the complex 
reflection amplitude $R=\sqrt(r)e^{i\theta}$ for a 1D disordered
sample, with  different lengths and with  different disorder
strengths. For our numerical solution, we have taken a fixed initial
reflection coefficient $r_0=.01$ for all realization of disorder as the 
Fokker-Planck equation for $P(r,\theta)$ is singular at $r=0$.
Gross statistical properties of the system are not expected to
change with this weak extra scatterer. It may, however, affect
the very sensitive details corresponding to the limit
$r\rightarrow 0$, $l\rightarrow 0$. Our numerical work is a
systematic study to observe the contribution of the phase
fluctuations to different averages.\\ 
On the basis of the results obtained, our conclusions are the
following: \\ 

{\bf (A) $P(r,\theta)$ and $P(\theta)$ distributions}.\\
1. The probability distribution $P(r,\theta)$  obeys
 two-parameter scaling to include the phase fluctuations and
merge to one-parameter scaling in the limit of weak disorder.
Full distribution of $P(r,\theta)$ can be described by the
length $l$ and disorder parameter $2k\xi$. \\
2. Phase distributions $P(\theta)$ also have the same
two-parameter scaling dependence for arbitrary disorder
strengths; and uniform phase distribution in the limit of weak
disorder, i.e., one-parameter scaling. 

{\bf (B) Phase $\theta$ distribution $P(\theta)$ in
different regimes of disorder}.\\
 1. The random phase approximation (RPA) implying uniform
 phase distribution over $2 \pi$ is valid for 
the condition $\xi /\lambda > 1$ ($\xi$ is localization length
and $\lambda$ is incoming wavelength), that is, in the weak
disorder limit. Physically, this means that the wave has to 
undergo multiple reflections before it moves through one
localization length. $P(\theta)$ is independent of the disorder
in the weak disorder limit $(2k\xi >>1)$.\\
2. In the strong disorder regime the distribution of the phase is
perfectly symmetric in the interval from $0$ and $2 \pi$, centering
at $\theta=\pi$. The distribution is independent of the disorder
in the strong disorder limit  $(2k\xi<<1)$.\\
3. In the intermediate disorder regime, phase distribution
is asymmetric about $\theta=\pi$ point in the interval $0$ and $2\pi$ . 
Also, the distributions are strongly disorder dependent. 

{\bf (C) Reflection Coefficient $(r)$ distributions $P(r)$:}\\
The probability distribution for the reflection coefficient
peaks at $r=1$ for large lengths $l \equiv L/\xi >>1$ \\
Though probability distribution $P(\theta)$ for the phase varies
qualitatively with the variation of the disorder strength, 
the probability distribution for the reflection coefficient
$P(r)$ does not change qualitatively with the disorder strength
for a fixed length of the sample.

{\bf (D) Distribution of $\rho$, $g$, $ln(\rho)$, $ln(g)$}:\\
 For large lengths of the sample (larger than the localization length)
the probability distribution for  $\rho$/ $g$ obeys the log-normal
distribution (i.e., $ln(\rho)$/$ln(g)$ obeys the normal
distribution) for all disorder strengths. An asymmetric phase
distribution does not affect the nature of the probability
distribution of the resistance, or the conductance, which stays
essentially log-normal.

{\bf (E) The mean and the rms fluctuations of $\rho$ and $g$:}\\
 1.The average resistance increases exponentially with the
         length of the sample.\\
 2.The average conductance decreases exponentially with the
         sample length. \\
3. The rms fluctuations are more than the average for both, the
   resistance and the conductance.

{\bf (F) The One-Parameter Scaling theory and the phase
}\\ 
 The one-parameter scaling theory is approximately valid for
the resistance and the conductance with $ln(\rho)\,\, or \,\,ln(g)$
as the correct scaling variable for large sample lengths.\\

\centerline{\bf ACKNOWLEDGEMENTS }
\vspace{.5cm}
I gratefully acknowledge N. Kumar for encouragement, stimulating
discussions, several suggestions and critical comments
throughout this work. I also thank T. V. Ramakrishnan and A. M.
Jayannavar for discussions and several suggestions. Thank to
Council for Scientific and Industrial Reseacsh (CSIR) for a
research fellowship and Super Computer Education and Research
Center(SERC), Indian Institute of Science  for computation
facilities. 

\bibliographystyle{alpha}

%
%
\begin{figure}
\caption{ Initial probability distribution, i.e., when the sample length
  $l=0$:
 (a) Probability distribution $P(r,\theta)$. 
 (b) Marginal probability distribution $P(r)$.
 (c) Marginal probability distribution $P(\theta)$. }
\label{pint}
\end{figure}
\begin{figure}
\caption{ Evolution of the probability $P(r,\theta)$ with the sample
length $l$ in the weak disorder regime for a fixed disorder
strength parameter $2k\xi=100$ .
Plots are for sample lengths: (a) $l=1$, (b) $l=2$, 
(c) $l=5$, and (d) $l=10$.}
\label{pevw}
\end{figure}
\begin{figure}
\caption{ Evolution of the probability $P(r,\theta)$ with the sample
length $l$ in the medium disorder regime for a fixed disorder
strength parameter $2k\xi=1$ .
Plots are for sample lengths: (a) $l=1$, (b) $l=2$, 
(c) $l=5$, and (d) $l=10$.}
\label{pevm}
\end{figure}
\begin{figure}
\caption{ Evolution of the probability $P(r,\theta)$ with the sample
length $l$ in the strong disorder regime for a fixed disorder
strength parameter $2k\xi=.001$.
Plots are for sample lengths: (a) $l=1$, (b) $l=2$, 
(c) $l=5$, and (d) $l=10$.}
\label{pevs}
\end{figure}
\begin{figure}
\caption{ Marginal probability distribution $P(r)$ and $P(\theta)$
separately against the sample length $l$. Three plots are for the fixed
disorder strength parameters: (a) Weak disorder, $2k\xi=100$, (b) Medium
disorder, $2k\xi=1$, and (c) Strong disorder, $2k\xi=.001$
(corresponding to Figs.2, 3 and 4). }
\label{marrt}
\end{figure}
\begin{figure}
\caption{ Phase $\theta$ distribution $P(\theta)$  against the disorder
parameter $2k\xi$ for a fixed sample length $l=1$.}
\label{ptl1}
\end{figure}
\begin{figure}
\caption{ Phase $\theta$ distribution $P(\theta)$ against the
disorder parameter $2k\xi$ for the asymptotic limit of large length
$l\rightarrow \infty$.}
\label{ptl2}
\end{figure}
\begin{figure}
\caption {Reflection coefficient $r$ distribution $P(r)$ against the
disorder parameter $2k\xi$ for a fixed sample length 
$l=1$.}
\label{prl1}
\end{figure}
\begin{figure}
\caption{ The resistance $\rho$ distribution $P(\rho)$ against
the disorder strength parameter $2k\xi$ for a fixed sample
length $l=10$.}
\label{prho}
\end{figure}
\begin{figure}
\caption{ The conductance $g$ distribution $P(g)$ against
the disorder strength parameter $2k\xi$ for a fixed sample
length $l=10$.}
\label{pcon}
\end{figure}
\begin{figure}
\caption{ Logarithm of the resistance/conductance $ln\rho/lng$
distribution $P(ln\rho)/P(lng)$  against the disorder 
strength parameter $2k\xi$ for a fixed sample length
$l=10$.} 
\label{pln}
\end{figure}
\begin{figure}
\caption{ Logarithm of the resistance/conductance $ln\rho/ln g$
distribution $P(ln\rho)/P(lng)$  against the sample
length $l$ in the medium disorder regime for a fixed disorder
strength parameter $2k\xi=1$.}
\label{plnmd}
\end{figure}
\begin{figure}
\caption{ Plot of the average and the rms fluctuation versus length $l$  for
the different disorder strength parameters $2k\xi$ for
       (a)  Reflection coefficient $r$,
       (b)  Resistance  $\rho$ and
       (c)  Conductance $g$. }
\label{average}
\end{figure}
\begin{figure}
\caption{ Plot of the  $var$ $ \,ln\rho$  \,\, vs \,\, 
 $<ln\rho>$ for different disorder strengths $2k\xi$.}
\label{pone}
\end{figure}
\end{document}